\documentclass[runningheads]{llncs}
\usepackage[T1]{fontenc}
\usepackage{graphicx,verbatim}
\usepackage{amsmath}
\usepackage{booktabs}
\usepackage{hyperref}
\usepackage{tabularx}
\begin{document}
\title{Rule-based Key-Point Extraction for MR-Guided Biomechanical Digital Twins of the Spine}
\titlerunning{Rule-based Key-Point Extraction in MRI for the Spine}
\author{
Robert Graf\inst{1,2}\orcidID{0000-0001-6656-3680} \and
Tanja Lerchl\inst{1}\orcidID{0000-0002-7888-3830} \and
Kati Nispel\inst{1}\orcidID{0000-0002-3833-5696} \and
Hendrik Möller\inst{1,2}\orcidID{0009-0001-1978-5894} \and
Matan Atad\inst{1,2}\orcidID{0000-0001-6952-517X} \and
Julian McGinnis\inst{2,4}\orcidID{0009-0000-2224-7600} \and
Julius Maria Watrinet\inst{3}\orcidID{0000-0003-0578-4142} \and
Johannes Paetzold\inst{5}\orcidID{0000-0002-4844-6955} \and
Daniel Rueckert\inst{2,6}\orcidID{0000-0002-5683-5889} \and
Jan S. Kirschke\inst{1}\orcidID{0000-0002-7557-0003}
}
\authorrunning{R. Graf et al.}
\institute{
Department of Diagnostic and Interventional Neuroradiology, School of Medicine, 
TUM University Hospital, Technical University of Munich, München, Germany \and
Institut für KI und Informatik in der Medizin, TUM University Hospital, 
Technical University of Munich, München, Germany \and
Sports Orthopedics Department, Klinikum Rechts der Isar, Technical University of Munich, München, Germany \and
Department of Neurology, School of Medicine, Technical University of Munich, Munich, Germany\and
Department of Radiology, Weill Cornell Medicine, New York, USA \and
Department of Computing, Imperial College London, London, UK
}
    
\maketitle              
\begin{abstract}
Digital twins offer a powerful framework for subject-specific simulation and clinical decision support, yet their development often hinges on accurate, individualized anatomical modeling. In this work, we present a rule-based approach for subpixel-accurate key-point extraction from MRI, adapted from prior CT-based methods. Our approach incorporates robust image alignment and vertebra-specific orientation estimation to generate anatomically meaningful landmarks that serve as boundary conditions and force application points, like muscle and ligament insertions in biomechanical models. These models enable the simulation of spinal mechanics considering the subject's individual anatomy, and thus support the development of tailored approaches in clinical diagnostics and treatment planning. By leveraging MR imaging, our method is radiation-free and well-suited for large-scale studies and use in underrepresented populations. This work contributes to the digital twin ecosystem by bridging the gap between precise medical image analysis with biomechanical simulation, and aligns with key themes in personalized modeling for healthcare.


\end{abstract}
\begin{figure}[htbp]
    \centering
    \includegraphics[width=\textwidth]{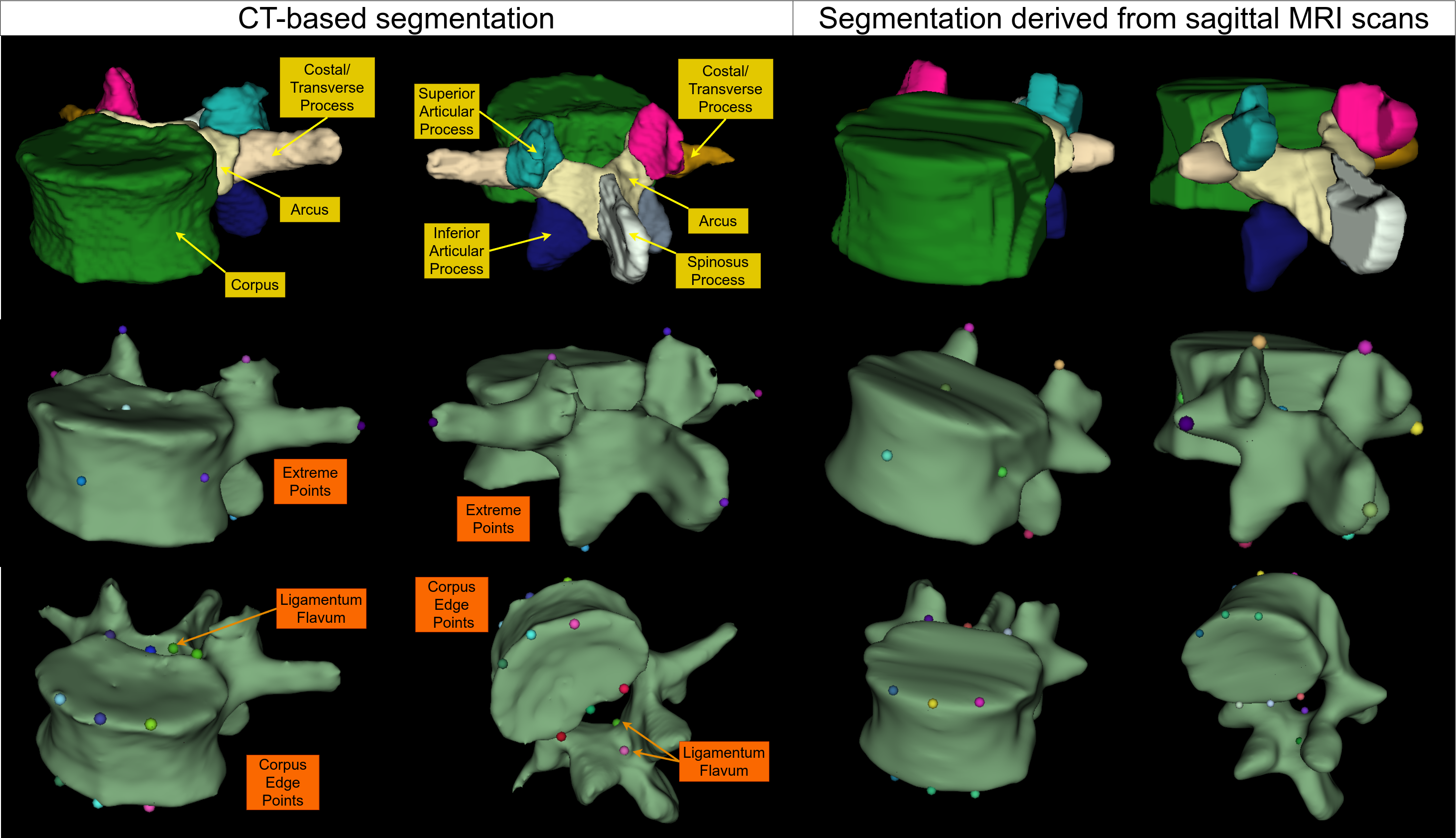}
    \caption{
    Example of two lumbar vertebrae. The left example is derived from 1 mm isotropic CT, the right from sagittal MRI with a resolution of 3.3 mm in the left–right direction. Top row: Subregion of the vertebra used for analysis. Middle row: Extreme points. Bottom row: Corpus edge and ligamentum flavum points.}
    \label{fig:landmark_examples}
\end{figure}
\label{raycast}

\section{Introduction}

Biomechanical modeling plays a critical role in understanding the mechanical behavior of the human spine and in studying musculoskeletal disorders. Finite element methods (FEM) are commonly used to assess local tissue stresses and deformations, supporting research into intervertebral disc (IVD) degeneration, scoliosis, and implant optimization \cite{balasubramanian2022development,el2020development,couvertier2017biomechanical,nispel2023recent}. Multibody systems (MBSs), on the other hand, capture the kinematics and dynamics of rigid body segments like vertebrae and are used to simulate spinal posture, joint loading, and musculoskeletal motion \cite{christophy2013modeling,lerchl2023multibody,wren2017biomechanical}. A core requirement for MBSs is the accurate identification of points of interest (POIs) on bones, which define joint axes, force application sites, and coordinate frames \cite{huynh2015development,lerchl2022validation}. Lerchl et al. \cite{lerchl2022validation} introduced a rule-based framework for extracting such POIs from CT scans, achieving voxel precision of attachments of ligaments and muscles. This enabled subject-specific modeling of spinal mechanics. However, CT-based modeling is limited in determining soft tissues such as IVDs. To address this, we adapt and extend this method for use with MRI, which offers superior soft tissue contrast but poses challenges due to its lower resolution in sagittal T2-weighted sequences and anisotropic voxels, making traditional pixel-based algorithms less reliable. Moreover, this existing approach often assumes vertebrae are aligned with image axes, a problematic assumption in patients with spinal deformities like scoliosis. We overcome this by computing vertebral orientations directly from the image, enabling robust modeling even in rotated or misaligned scans. To promote reproducibility and further research, we release our implementation as an open-source script, facilitating MRI-based MBM workflows and expanding biomechanical digital twin applications to settings where soft tissue characterization is essential.

\begin{figure}[htbp]
    \centering
    \includegraphics[width=\textwidth]{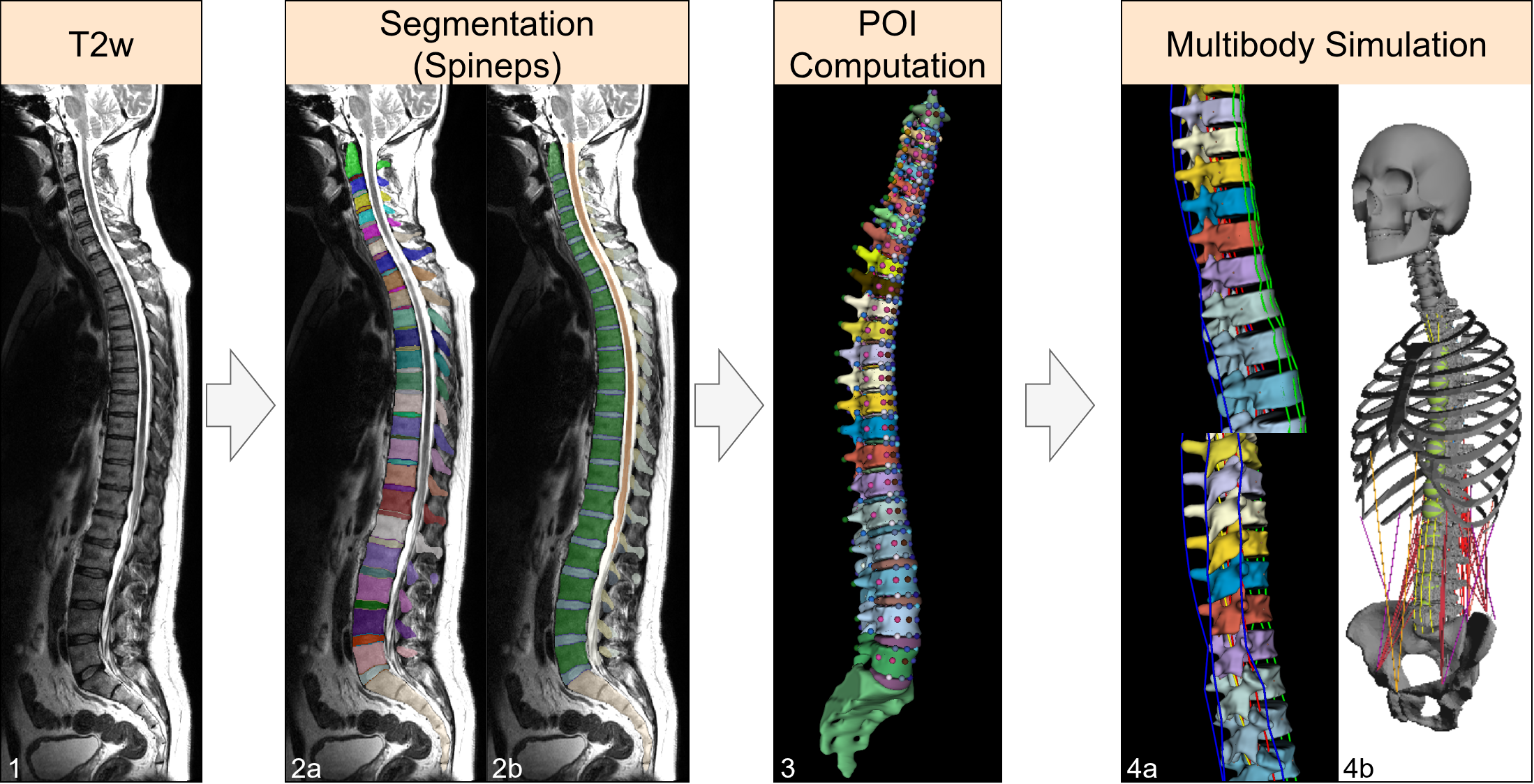} 
    \caption{
       Full pipeline for the generation of multibody models from MR imaging. A sagittal T2-weighted MRI (1) is segmented using Spineps, a machine learning based pipeline for automated whole spine segmentation of level-wise vertebrae and intervertebral discs (2a) as well as respective subregions (2b). Based on these segmentation masks, individual points of interest (POIs) are calculated (3) to define ligament (4a) and spinal muscle attachments to other regions (4b) in the subsequent multibody model of the torso.
    }
    \label{fig:full-pipline}
\end{figure}
Building on our approach to directly estimate vertebral orientation from image data, we situate our work within a broader landscape of anatomical landmark detection. Traditionally, POI prediction has focused on bone landmarks, such as those in the head \cite{lachinov2020cephalometric,o2018attaining,tao2023automatic} or lower limbs \cite{furmetz2021three,furmetz2019three}. Multi-stage prediction networks are commonly employed to refine these estimates across successive processing steps. Several methods have been developed to estimate vertebral orientation, anatomical lines, or discrete landmarks. For instance, Galbusera et al. \cite{galbusera2020estimating} used a ResNet-based regression model to directly estimate 3D vertebral axes from sagittal radiographs, achieving angular errors below $3^\circ$ in more than 86\% of cases. More recent graph-based strategies, such as Burgin et al. \cite{burgin2023robust}, detect pedicle and vertebral body landmarks using a U-Net, which are then processed via a graph neural network to infer vertebral pose and inter-point relationships. Other work has focused on line-based representations; Zhang et al. \cite{zhang2024dual} proposed a dual-coordinate model that reconstructs vertebral lines from sparse landmark inputs, improving resilience to anatomical variation and partial visibility. In surgical planning contexts, Zhang et al. \cite{zhang2023improving} introduced a YOLO-inspired network that jointly regresses vertebral translation and orientation as quaternions, achieving angular errors around $2.55^\circ$. Lastly, landmark-based methods like that of Khanal et al. \cite{khanal2019automatic} use vertebral corner regression to estimate vertebral tilt angles, which are especially relevant for scoliosis analysis. However, despite this progress, there remains no publicly available benchmark for vertebral orientation estimation, and most current approaches rely heavily on labor-intensive manual annotations.

In summary, we present an open-source framework for MRI-based extraction of vertebral orientation and points of interest, enabling more accurate multibody spinal modeling in the presence of soft tissue and anatomical variability. By relaxing alignment assumptions and integrating vertebral pose estimation directly from image data, our method broadens the applicability of musculoskeletal simulations, particularly in cases with spinal deformities.

\section{Method}

We employed the point-of-interest (POI) generation code developed in Lerchl et. al. \cite{lerchl2022validation} and evaluated the modifications required to adapt it for use with MRI. The method operates exclusively on segmentation and is independent of the imaging modality. Only the quality and resolution of the segmentation affect the algorithm. It depends on a detailed vertebra substructure segmentation, including the separation of anatomical subregions: vertebral body, arcus, spinous process, costal processes (left/right), superior articular processes (left/right), and inferior articular processes (left/right). For this purpose, we leveraged the open-access segmentation model SPINEPS \cite{graf2023denoising,moller_spinepsautomatic_2024}, which is capable of producing such fine-grained segmentations from sagittal T2-weighted MRI. However, clinical sagittal T2w scans typically suffer from low in-plane resolution in the left-right direction (3–4 mm), due to practical constraints like acquisition time and the anatomical extent of the spine. When applying the original pixel-based methods by Lerchl et al. \cite{lerchl2022validation}, we observed substantial inaccuracies under these conditions. Additionally, the method assumes that the vertebrae are aligned with the volume’s left-right axis — a condition that may not hold in cases involving spinal deformities such as scoliosis. To improve robustness and anatomical accuracy, we introduce an algorithm that estimates the local orientation of each vertebra, decoupling the landmark computation from both voxel spacing and orientation. This approach replaces voxel-based assumptions with sub-pixel-accurate geometric logic, enabling consistent POI definition across varied scan orientations and resolutions. We release the full implementation under the Python package \href{https://github.com/Hendrik-code/TPTBox}{TPTBox}, including tools to recompute POIs in alternative coordinate systems—such as different voxel spacings, ITK world space, or NIfTI world space—to support reproducibility and integration into broader workflows.

\subsection{Vertebra Orientation}
\begin{figure}[tp]
    \centering
    \includegraphics[width=\textwidth]{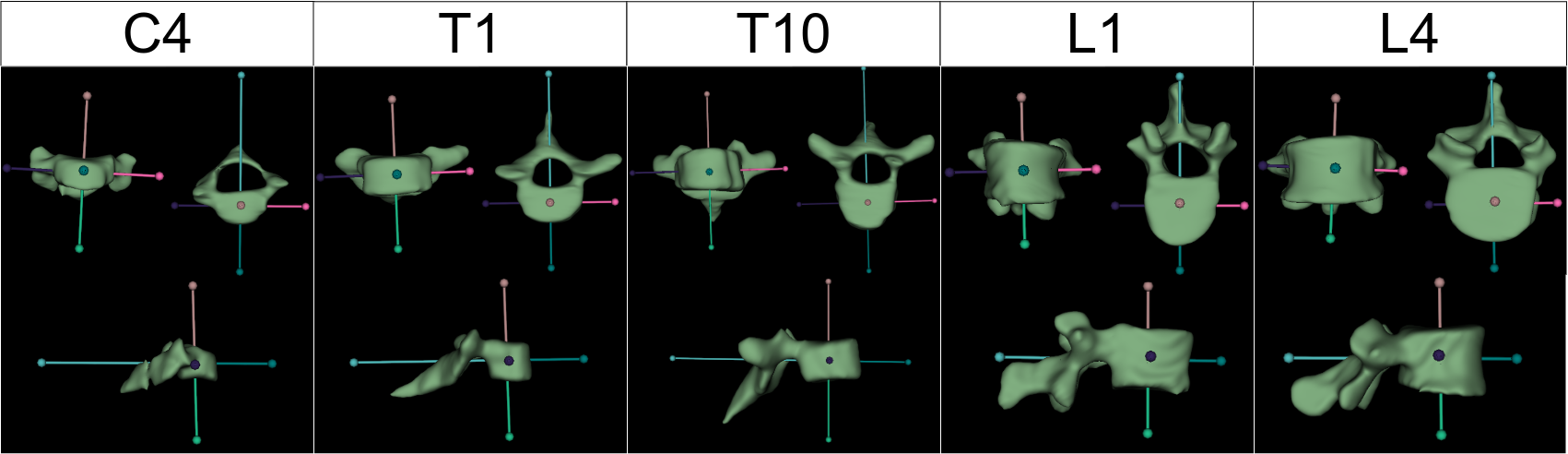} 
    \caption{
        Front, top, and right views of  randomly selected vertebrae, visualized with the computed local coordinate system overlaid as directional whiskers. The top/bottom (cranio-caudal) axis is defined relative to adjacent vertebral bodies using a spline through their centers of mass. Due to anatomical asymmetries in structures such as the processus spinosus and arcus vertebrae, determining the front/back direction can be challenging. In particular, the L4 vertebra shown here exhibits significant asymmetry, which results in a slight rotation of the computed front/back direction compared to the visually expected anatomical posterior.
    }
    \label{fig:directions}
\end{figure}
\label{raycast}
The original implementation used the cardinal directions of the image (left/right and front/back) as proxies for vertebral orientation. While this assumption holds for most healthy spines, it can break down in emergency settings, cases where patients cannot maintain a standard posture, or in the presence of spinal deformities such as scoliosis. We kept the original approach for computing the superior-inferior (up/down) direction: a spline is fitted through the centers of mass of the vertebral bodies, and the first derivative of this spline defines the local up/down direction. This approach avoids errors introduced by assuming that the vertebral body is cuboidal or that the endplates are flat and parallel.

To compute the second anatomical direction, we extract the masks of the spinous process and the arcus vertebrae. These structures are projected onto a plane orthogonal to the up/down direction. The geometric centers of the projected masks are then computed, and a vector connecting the computed center with the center of mass of the vertebral corpus defines the second direction. The two vectors define a plane, and we recompute the front/back vector to be orthogonal to the up/down vector. The third direction is obtained as the cross product of the first (up/down) and second (front/back) directions, thereby forming an orthonormal local vertebral coordinate system.

\subsection{Ray Casts}
\begin{figure}[htbp]
    \centering
    \includegraphics[width=\textwidth]{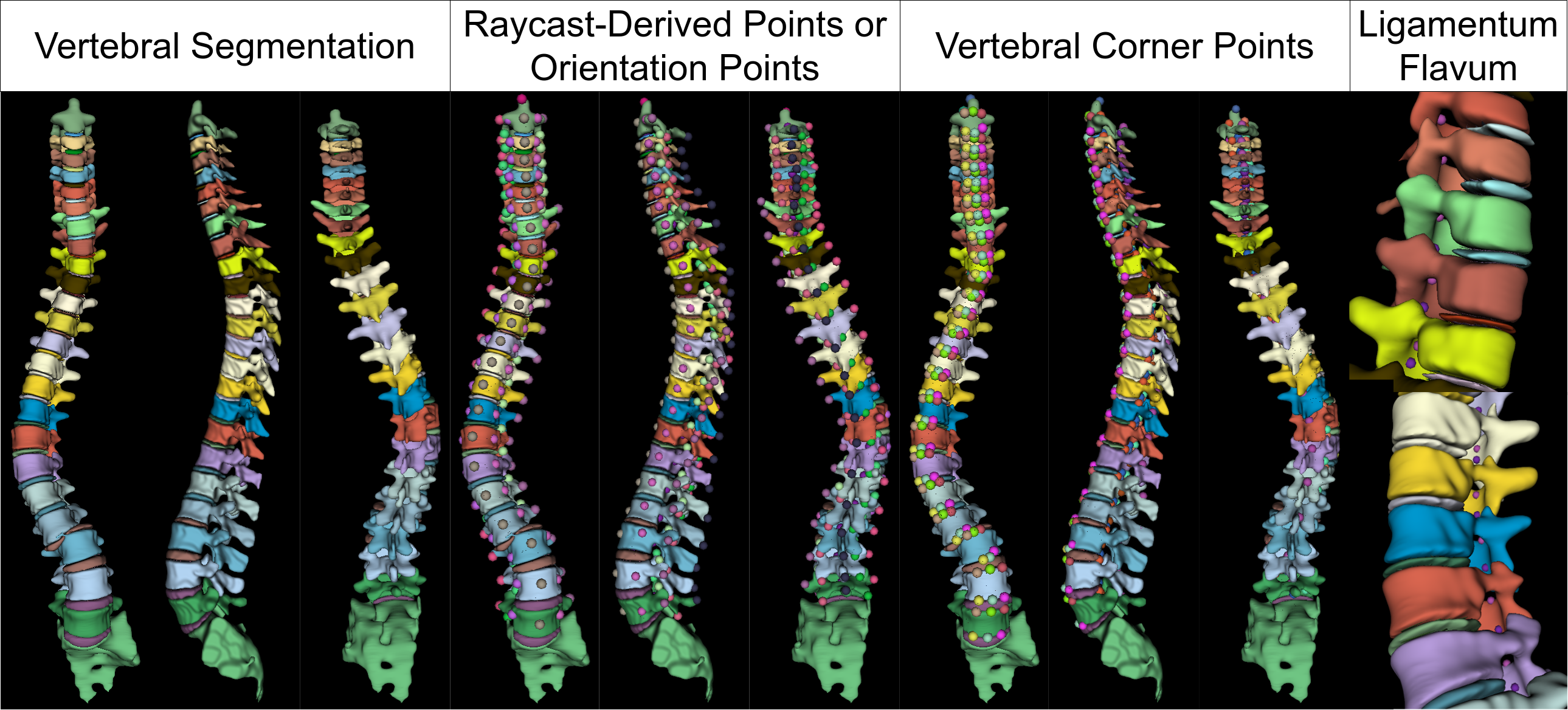} 
    \caption{
        Visualization of vertebral landmark extraction on a previously unseen sagittal T2-weighted MRI scan of a scoliotic spine. The first three panels show the vertebra segmentation overlaid with points, each with frontal, left, and back views. From left to right: segmentation only, the computed raycasting-based points, and corner landmarks. These highlight the robustness of the method under spinal curvature and low in-plane resolution. The final panel provides a close-up axial view of the spinal canal, focusing on the \emph{ligamentum flavum} landmarks, which are accurately positioned on the posterior vertebral arch despite anatomical variability.
    }
    \label{fig:landmark_extraction}
\end{figure}
\label{raycast}
The endpoints of all vertebral processes are now computed via raycasting from the center of mass of each corresponding subregion, using the previously computed local vertebral coordinate system. The superior and inferior articular processes are defined along the superior-inferior axis (up and down directions, respectively). For the transverse processes, the raycasting direction is given by the vector

$$
\mathbf{a} = 0.5 \cdot \mathbf{l} + 0.5 \cdot \mathbf{p},
$$

where $\mathbf{l}$ is the lateral (left/right) direction and $\mathbf{p}$ is the posterior (backward) direction. For the spinous process, the raycasting direction is defined as

$$
\mathbf{a} = \mathbf{d} + 0.2 \cdot \mathbf{p},
$$

where $\mathbf{d}$ is the inferior (downward) direction and $\mathbf{p}$ is again the posterior direction. These direction vectors were empirically chosen to best match the observed anatomical trajectories and to ensure robust point localization despite inter-subject variability.

The vertebral body is assigned one surface point in each of the six cardinal directions by raycasting from its center of mass along the corresponding direction vectors of the local vertebral coordinate system.

\subsection{Vertebral corpus corners with Sub-Voxel Bisection}

In the original method, vertebral body corner points were extracted by selecting the sagittal slice intersecting the center of mass of the vertebral corpus. A Sobel filter was applied to detect edge candidates, and an image-aligned 2D bounding box was drawn around the vertebral body. The closest candidate points to the corners of this bounding box were then selected as corner landmarks. Additional intermediate edge points were computed and projected onto the vertebral surface.

In our updated approach, we computed the vertebral body midpoints previously via raycasting (Section \ref{raycast}). To compute the vertebral corpus corner points, we employ a 2D bisection search initialized at the center of mass and oriented along the local up/down and front/back direction vectors. The search iteratively steps outward in each direction and, upon exiting the segmentation boundary, the step size is halved repeatedly until a predefined precision threshold is reached. We interpolate at the tested coordinate to allow for subvoxel-accurate landmark placement that is robust to anatomical variability.

We compute the two additional anatomical landmarks corresponding to the \textit{ligamentum flavum}, located in the same axial plane as the vertebral body corners but positioned on the anterior surface of the vertebral arch. The same 2D bisection strategy is used, starting from the center of mass of the arcus and constraining the search domain to the arcus segmentation mask rather than the corpus.

\subsection{Shifted points}
The vertebral corners and cardinal direction landmarks are also computed using an offset in the left and right directions of the local vertebral coordinate system. These landmarks are anatomically motivated by the attachment of the anterior longitudinal ligament to the vertebral bodies. To improve anatomical accuracy, especially in the upper thoracic and cervical spine, we refined the original heuristic. Previously, the lateral shift was defined as one-third of the distance between the centers of mass of the superior articular processes. We now scale this shift using a vertebra-dependent factor

$$
f = \frac{12 - v_{id}}{11} + 1 \quad \text{for} \quad v_{id} \leq 11,
$$

where $v_{id}$ is the vertebra index counted from the top, with C1 assigned as 1. This scaling accounts for the stronger shrinking of the vertebral bodies compared to the posterior structure in the neck and upper thoracic region.

\section{Experiments and Results}

\subsection{Vertebra Orientation}

To evaluate our new rotation estimation method, we randomly selected 90 vertebrae (from 20 subjects; 11 Female) from the VerSe2020 challenge dataset \cite{sekuboyina2020verse}. We manually measured the angular deviation between the estimated and true posterior directions. We report the mean angular deviation in degrees ($^\circ$) and provide the fraction of results falling below two thresholds: one indicating excellent results and the other indicating catastrophic failures. A rotational deviation of $\leq 3^\circ$ is rarely noticeable in qualitative inspection, so we use this as a practical threshold for an “excellent” orientation estimate. Conversely, deviations exceeding $10^\circ$ were considered “catastrophic.” The comparison in Table~\ref{table:1} highlights how often each method meets these criteria.

\newcolumntype{C}[0]{>{\centering\arraybackslash}X}
\newcolumntype{L}[0]{>{\arraybackslash}X}
\newcolumntype{R}[0]{>{\raggedright\arraybackslash}X}
\newcolumntype{U}[1]{>{\arraybackslash}m{#1}}

\begin{table}[h!]

\centering
\caption{Comparison of our proposed backward direction computation to a naive 3D center-of-mass (CMS) approach. Angles are measured in degrees ($^\circ$). We report the mean $\pm$ standard deviation and the fraction of results with angular error below $3^\circ$ and $10^\circ$.}
\begin{tabularx}{\linewidth}{@{}U{4.8cm}CCC@{}}
\toprule
 & {$^\circ$ Mean $\pm$ Std $\downarrow$} & {Fraction $\leq3^\circ$ $\uparrow$} & {Fraction $\leq10^\circ$ $\uparrow$} \\
\midrule
3D CMS (all posterior structures) & 5.78$\pm$10.03 &0.39 (35/90) &0.91 (82/90) \\
3D CMS (Arcus and Spinosus)       & 2.87$\pm$6.84  &0.70 (63/90) &0.99 (89/90) \\
\textbf{2D Projection (ours)}     & \textbf{1.72$\pm$1.76} & \textbf{0.80 (72/90)} & \textbf{1.00 (90/90)} \\
\bottomrule
\label{table:1}
\end{tabularx}
\end{table}

The original implementation did not include orientation estimation. We therefore experimented with several strategies that derive the vertebral coordinate system from the 3D centers of mass (CMS) of automatically extracted subregions. Owing to pronounced structural asymmetries and inter-subject anatomical variability, a naive 3D CMS of all posterior structures produced a front/back (anterior–posterior) direction that was off by 5.78 ± 10.03$^\circ$ on average and exceeded 10$^\circ$ in 9\% of the cases (8/90). This failure mode occurred almost exclusively in cervical levels, where the posterior elements are markedly skewed. Restricting the CMS to the arcus and spinosus parts mitigated this problem (2.87 ± 6.84$^\circ$, with only 1/90 cases $> 10^\circ$), yet the spread was still larger than we considered acceptable. To further stabilize the estimate, we introduced a regularization step: all relevant posterior voxels are first projected onto a 2D plane orthogonal to the superior–inferior axis, after which the 2D CMS is computed and re-embedded in 3D. This simple projection removes most of the out-of-plane asymmetry and shrinks the error to 1.72 ± 1.76$^\circ$. Crucially, the method now achieves an error below 3$^\circ$ in 80\% (72/90) and below 10$^\circ$ in 100\% (90/90) of the vertebrae, which we deem sufficiently accurate for downstream shape analysis and visualization.

In summary, naively averaging all posterior voxels is prone to large angular errors, particularly in the cervical spine, whereas our 2D-projection strategy delivers robust, sub-3$^\circ$ accuracy in four out of five vertebrae and never exceeds 10$^\circ$.

\subsection{Points for Multi-Body Simulation}

To evaluate the reliability of our method for use in MBS, we tested it on 37 full spine segmentations. Two experts with 4 and 7 years of experience in Spine CT and MRI imaging assessed its performance. Failures occurred only in cases where the underlying segmentation was flawed.

We validated our point placement using an existing MBS framework \cite{lerchl2022validation}. Despite operating at a lower resolution, we observed no large discrepancy for straight spines, compared to existing CT-based point extraction. However, in some cases, we noticed large forces between the corner points of adjacent vertebral bodies. Upon investigation, we determined that this occurred when the vertebral bodies were closely aligned in the up/down direction but offset in the front/back or left/right direction. This issue arises from the definition of the frontal ligament used in the simulation, not from inaccuracies in the corner point placement. While this might indicate real tension in the anterior ligament, it is more likely due to the real anatomical attachment point located closer to the vertebral center. Accurately extracting the ligament path would be necessary to resolve this ambiguity, but this is not feasible with CT and is currently not available in MRI.

\section{Discussion and Conclusion}

We presented an advanced version of a rule-based POI extraction pipeline for the spine. The pipeline now supports MRI input, even with low left-right resolution, and can compensate for relative vertebral rotation—an essential capability for analyzing scoliotic spines. The used segmentation network and trained weights are publicly available, along with our enhanced point computation method. The entire POI computation completes in under a minute on a single CPU thread for a whole spine. Additionally, we provide tools for saving and loading the computed points and resampling them to different coordinate systems, such as voxel space, ITK, and NIfTI global space. The points can also be exported in the "mkr.json" format, allowing for easy import, editing, and visualization in 3D Slicer.

Our generated points provide a solid foundation for further development. While rule-based systems are effective, they tend to accumulate exceptions and special cases, such as fractured vertebrae, vertically misaligned segments, or the presence of metal implants, which become difficult to handle manually. In such scenarios, it is more efficient to correct the rule-based outputs and allow a deep learning model to generalize from them. Starting from our initial point annotations, it should be feasible to generate datasets that can be refined, corrected, and expanded, paving the way for robust, learning-based point prediction pipelines.


\appendix
\subsubsection{\ackname}
The research for this article received funding from the European Research Council (ERC) under the European Union’s Horizon 2020 research and innovation program (101045128—iBack-epic—ERC2021-COG).

Our Code is available in the Python package \href{https://github.com/Hendrik-code/TPTBox}{https://github.com/Hendrik-code/TPTBox}

\subsubsection{\discintname} The authors have no competing interests to declare that are
relevant to the content of this article. 

\bibliographystyle{splncs04}
\bibliography{bib}
\end{document}